\documentclass[aps,reprint,prl,superscriptaddress,showpacs,floats,floatfix]{revtex4-2}
\usepackage{morefloats}
\usepackage{mathptmx}
\usepackage{indentfirst}
\usepackage{amsmath,graphicx,dcolumn}
\usepackage[colorlinks,citecolor=green]{hyperref}
\usepackage[usenames]{color}
\usepackage{datetime}
\usepackage{textcomp}
\usepackage{ulem}
\usepackage{float}
\usepackage{titlesec}
\titleformat{\section}{\large\bfseries\raggedright}{\thesection}{1em}{}

\newcommand{\blue}{\textcolor{blue}}

\newcommand*{\LNO}{La$_4$Ni$_3$O$_{10}$}
\newcommand*{\mjcm}{$\mu$J/cm$^2$}
\begin{document}

\title{Electronic Nematicity Revealed by Polarized Ultrafast Spectroscopy in Bilayer La$_3$Ni$_2$O$_7$}

\author{Qi-Yi Wu}
\thanks{These authors contributed equally to this work}
\affiliation{Hunan Key Laboratory of Super Microstructure and Ultrafast Process, School of Physics, Central South University, Changsha 410012, Hunan, China}
\affiliation{School of Materials Science and Engineering, Central South University, Changsha 410083, Hunan, China}

\author{De-Yuan Hu}
\thanks{These authors contributed equally to this work}
\affiliation{Institute of Neutron Science and Technology, Guangdong Provincial Key Laboratory of Magnetoelectric Physics and Devices, School of Physics, Sun Yat-Sen University, Guangzhou 510275, Guangdong, China}

\author{Chen Zhang}
\affiliation{Hunan Key Laboratory of Super Microstructure and Ultrafast Process, School of Physics, Central South University, Changsha 410012, Hunan, China}

\author{Hao Liu}
\affiliation{Hunan Key Laboratory of Super Microstructure and Ultrafast Process, School of Physics, Central South University, Changsha 410012, Hunan, China}

\author{Bo Chen}
\affiliation{Hunan Key Laboratory of Super Microstructure and Ultrafast Process, School of Physics, Central South University, Changsha 410012, Hunan, China}

\author{Ying Zhou}
\affiliation{Hunan Key Laboratory of Super Microstructure and Ultrafast Process, School of Physics, Central South University, Changsha 410012, Hunan, China}

\author{Zhong-Tuo Fu}
\affiliation{Hunan Key Laboratory of Super Microstructure and Ultrafast Process, School of Physics, Central South University, Changsha 410012, Hunan, China}

\author{Chun-Hui Lv}
\affiliation{Hunan Key Laboratory of Super Microstructure and Ultrafast Process, School of Physics, Central South University, Changsha 410012, Hunan, China}

\author{Zi-Jie Xu}
\affiliation{Hunan Key Laboratory of Super Microstructure and Ultrafast Process, School of Physics, Central South University, Changsha 410012, Hunan, China}

\author{Hai-Long Deng}
\affiliation{Hunan Key Laboratory of Super Microstructure and Ultrafast Process, School of Physics, Central South University, Changsha 410012, Hunan, China}

\author{Meng-Wu Huo}
\affiliation{Institute of Neutron Science and Technology, Guangdong Provincial Key Laboratory of Magnetoelectric Physics and Devices, School of Physics, Sun Yat-Sen University, Guangzhou 510275, Guangdong, China}

\author{H. Y. Liu}
\affiliation{Beijing Academy of Quantum Information Sciences, Beijing 100085, China}

\author{Jun Liu}
\affiliation{School of Materials Science and Engineering, Central South University, Changsha 410083, Hunan, China}

\author{Yu-Xia Duan}
\affiliation{Hunan Key Laboratory of Super Microstructure and Ultrafast Process, School of Physics, Central South University, Changsha 410012, Hunan, China}

\author{Dao-Xin Yao }
\affiliation{Institute of Neutron Science and Technology, Guangdong Provincial Key Laboratory of Magnetoelectric Physics and Devices, School of Physics, Sun Yat-Sen University, Guangzhou 510275, Guangdong, China}

\author{Meng Wang}
\email{Corresponding author: wangmeng5@mail.sysu.edu.cn}
\affiliation{Institute of Neutron Science and Technology, Guangdong Provincial Key Laboratory of Magnetoelectric Physics and Devices, School of Physics, Sun Yat-Sen University, Guangzhou 510275, Guangdong, China}

\author{Jian-Qiao Meng}
\email{Corresponding author: jqmeng@csu.edu.cn}
\affiliation{Hunan Key Laboratory of Super Microstructure and Ultrafast Process, School of Physics, Central South University, Changsha 410012, Hunan, China}

\date{\today}

\begin{abstract}

We report a polarized ultrafast pump-probe study of the normal-state electronic dynamics in bilayer La$_3$Ni$_2$O$_7$ and trilayer La$_4$Ni$_3$O$_{10}$ single crystals at ambient pressure. While both nickelates exhibit density-wave (DW) transitions accompanied by the opening of a quasiparticle relaxation bottleneck, their electronic responses display strikingly different symmetry properties. La$_4$Ni$_3$O$_{10}$ maintains an isotropic optical response across the entire temperature range. In contrast, La$_3$Ni$_2$O$_7$ exhibits a pronounced twofold ($C_2$) anisotropy in its low-temperature electronic dynamics. This electronic nematicity, evident in both the relaxation dynamics and the polarization-dependent effective bottleneck energy scales, is strongly modified below 115 K, suggesting coupling or competition with a secondary DW-like instability reported by complementary probes. The presence of macroscopic electronic anisotropy in the bilayer system, and its absence in the trilayer system, suggests a possible relation between electronic nematic correlations and the superconducting normal state in La$_3$Ni$_2$O$_7$ that deserves further exploration.

\end{abstract}

\maketitle

The recent discovery of high-temperature superconductivity in bilayer La$_3$Ni$_2$O$_7$ \cite{HSun2023} and trilayer La$_4$Ni$_3$O$_{10}$ \cite{YZhu2024} under pressure has established a new high-temperature superconductor (HTSC) platform following cuprates and iron-based superconductors \cite{YNZhang2024, EKKo2025, GZhou2025, ZLiu2023, NWang2024, JZhan2025, QLi2024, PPuphal2024, JXWang2024, QGYang2024, EZhang2025, YGu2025, YChen2025}, motivating systematic comparisons to elucidate the underlying microscopic mechanisms. As shown in Figure \blue{1}(a), these nickelates belong to the Ruddlesden-Popper perovskite family ($n=2, 3$), featuring LaNiO$_3$ layers interleaved by a single rock-salt-type LaO layer. The central ingredients of La$_3$Ni$_2$O$_7$ and La$_4$Ni$_3$O$_{10}$ are the bilayer and trilayer NiO$_2$ planes, which are linked via shared apical oxygen, enabling stronger interlayer coupling \cite{JZhang2017, VVPoltavets2010, ZLuo2023}. The multi-orbital feature of electronic structure near the Fermi level resemble iron-based superconductors, primarily contributed by Ni $3d_{x^2-y^2}$ and $3d_{z^2}$ orbitals \cite{ZLuo2023, VChristiansson2023, FLechermann2023, YZhang2023, JYang2024, JLi2024, HLi2017, ZHuo2025A}.\setlength{\parskip}{0pt}

In HTSCs, superconductivity typically emerges when the parent antiferromagnetic (AFM) state and nematicity are suppressed by doping \cite{BKeimer2015, HHosono2015}. Similarly, superconductivity in La$_3$Ni$_2$O$_7$ and La$_4$Ni$_3$O$_{10}$ appears after the density wave (DW) order is suppressed by pressure. Elucidating the normal state is thus fundamental. In La$_3$Ni$_2$O$_7$, a spin density wave (SDW) order below $T_{\rm SDW}\sim$ 130-150 K is corroborated by various experiments \cite{KChen2024, XChen2024, IPlokhikh2025, TXie2024, rKhasanov2025, QYWu2025, YMeng2024, DZhao2025}. Notably, the $T_{\rm SDW}$ exhibits contrasting pressure dependence in different probes \cite{DZhao2025, rKhasanov2025, QYWu2025, YMeng2024}, suggesting two distinct DW transitions. La$_4$Ni$_3$O$_{10}$ exhibits a DW transition near $T_{\rm DW} \sim$ 135 K, confirmed by heat capacity, transport, NMR, and $\mu$SR measurements \cite{MZhang2025, DZhao2025, MKakoi2024, TFukamachi2001, tFukamachi2001, YCao2025, RKhasanov2025}. A DW gap opening is also observed \cite{HLi2017, ZLiu2024, SXu2025, sXu2025, YLi2025, CZhang2025, XDu2024}. X-ray and neutron diffraction studies indicate the coexistence of intertwined SDW and charge density wave (CDW) orders with layer-dependent features \cite{JZhang2020}. Unlike La$_3$Ni$_2$O$_7$, DW and SDW transition temperatures in La$_4$Ni$_3$O$_{10}$ are suppressed under pressure \cite{RKhasanov2025, sXu2025}. Despite significant progress, a comprehensive understanding of the normal state and its link to superconductivity remains elusive.\setlength{\parskip}{0pt}

\begin{figure*}[t]
\vspace*{-0.2cm}	\begin{center}
\includegraphics[width=1.7\columnwidth]{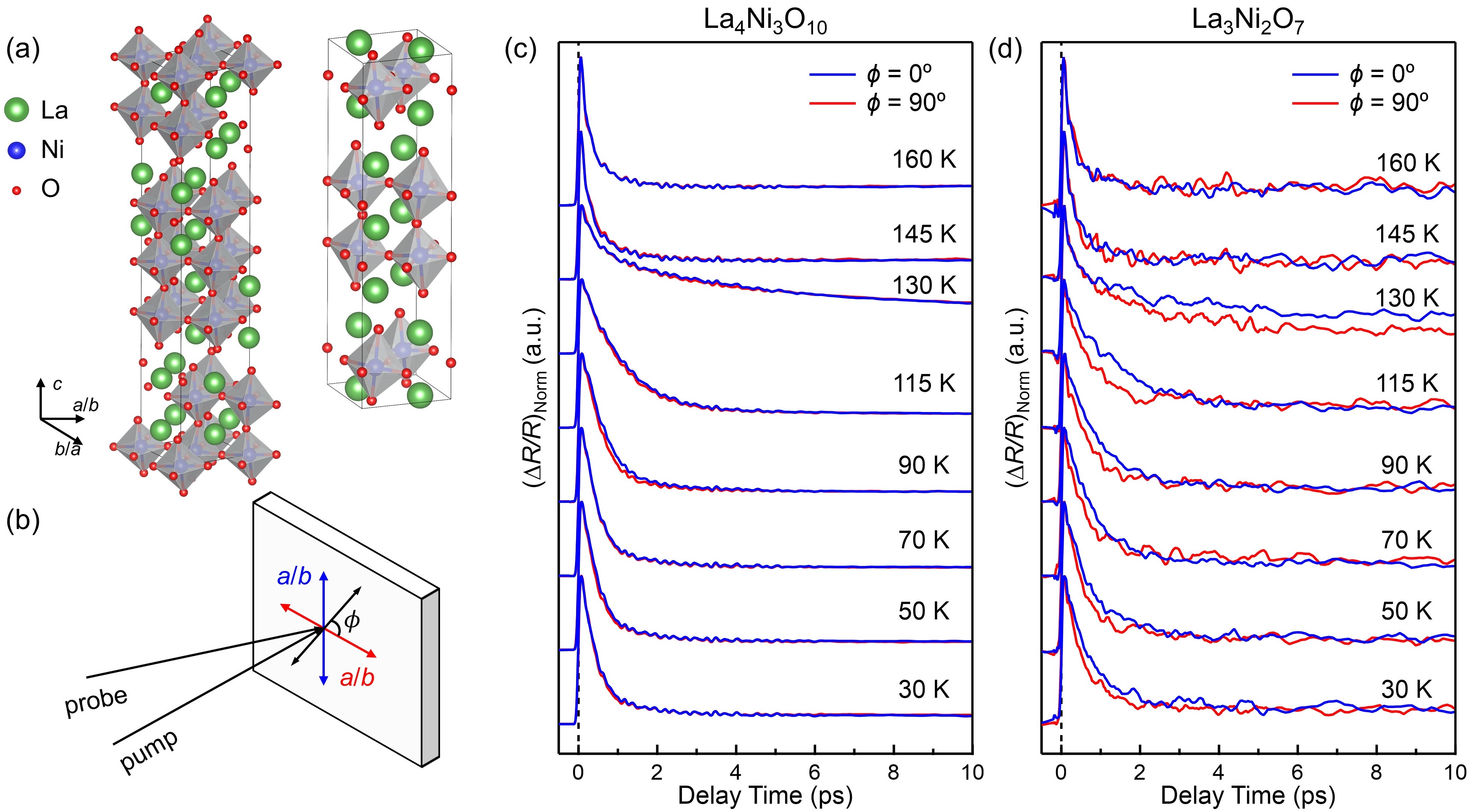}
\end{center}
\vspace*{-0.7cm}
\caption{(color online) Temperature-dependent normalized ultrafast reflectivity of La$_4$Ni$_3$O$_{10}$ and La$_3$Ni$_2$O$_7$ along $\phi$ = 0$^{\circ}$ and $\phi$ = 90$^{\circ}$. (a) Crystalline structure of La$_4$Ni$_3$O$_{10}$ and La$_3$Ni$_2$O$_7$. (b) The schematic diagram of the polarized pump-probe spectroscopy. (c) and (d) The normalized ($\Delta R$/$R$)$_{Norm}$ as a function of delay time at various temperatures for La$_4$Ni$_3$O$_{10}$ and La$_3$Ni$_2$O$_7$, probed with $\phi$ = 0$^{\circ}$ and $\phi$ = 90$^{\circ}$, respectively. The gray-shaded area in panel (d) indicates the delay time range where the $\Delta$R/R signals of  La$_3$Ni$_2$O$_7$ exhibit distinct difference along $\phi$ = 0$^{\circ}$ and $\phi$ = 90$^{\circ}$. Measurements were conducted at a pump fluence of $\sim$9.9 {\mjcm}.}
\label{Figure:1}
\end{figure*}

In this work, we conducted polarized ultrafast pump-probe spectroscopy on single crystals of bilayer La$_3$Ni$_2$O$_7$ and trilayer La$_4$Ni$_3$O$_{10}$ under ambient pressure. Comparing the normalized transient reflectivity along $\phi$ = 0$^{\circ}$ and $\phi$ = 90$^{\circ}$ revealed a distinct contrast between the two compounds. The normalized transient reflectivity of La$_4$Ni$_3$O$_{10}$ remains isotropic across the entire temperature range investigated. In contrast, La$_3$Ni$_2$O$_7$ shows isotropic behavior at high temperatures, while a pronounced anisotropy becomes apparent upon cooling to lower temperatures. Further polarization-dependent measurements on La$_3$Ni$_2$O$_7$ demonstrated that rotational symmetry is broken from isotropic to a two-fold rotational symmetry in the temperature regime where the density-wave gap is well resolved. This rotational symmetry breaking, being present in La$_3$Ni$_2$O$_7$ and absent in La$_4$Ni$_3$O$_{10}$, suggests that electronic anisotropy may play a role in enhancing superconductivity in nickelate superconductors. These findings provide critical information for understanding the mechanism of superconductivity in these materials.

The single crystals of La$_3$Ni$_2$O$_7$ and La$_4$Ni$_3$O$_{10}$ with well-defined (001) cleavage planes were grown using a vertical optical image floating-zone method \cite{HSun2023, JLi2024}. Ultrafast pump-probe measurements were performed in the temperature range of 10-200 K, with all data collected at  a low pump fluence of $\sim$ 9.9 {\mjcm}. In the polarization-dependent experiments, the polarization of the pump and probe beams was adjusted by rotating a half-wave plate and a polarizer placed in front of the sample. The polarization angle ($\phi$) is defined as the rotated angle of polarization of the probe pulses relative to the horizontal $p$-polarization, as shown in Figure \blue{1}(b). Detailed descriptions of all measurement procedures can be found in Section I of Supporting Information \cite{CZhang2022, QYWu2023}.\setlength{\parskip}{0pt}

Figures \blue{1}(c) and \blue{1}(d) present the normalized transient reflectivity ($\Delta R$/$R$)$_{\rm Norm}$ of La$_4$Ni$_3$O$_{10}$ and La$_3$Ni$_2$O$_7$ measured along $\phi$ = 0$^{\circ}$ and $\phi$ = 90$^{\circ}$ at different temperatures, respectively. Both nickelates exhibit characteristic anomalies near their respective density-wave (DW) transition temperatures ($T_{\rm SDW}$ = 140 K for La$_3$Ni$_2$O$_7$ \cite{QYWu2025}, $T_{\rm DW}$ = 135 K for La$_4$Ni$_3$O$_{10}$ \cite{sXu2025,SXu2025}), signaling the opening of a DW energy gap. For La$_4$Ni$_3$O$_{10}$, the ($\Delta R$/$R$)$_{\rm Norm}$ curves show complete overlap across all temperatures for both polarizations, demonstrating an optically isotropic response, despite the material's lower monoclinic crystal symmetry. In sharp contrast, La$_3$Ni$_2$O$_7$ develops a distinct optical anisotropy that is clearly established within the SDW state below $T_{\rm SDW}$. While the high-temperature response remains isotropic, cooling to 130 K induces a clear deviation for $t_{delay}>0.8$ ps. This onset temperature coincides with observations of transient ellipticity anisotropy \cite{QYWu2025}. This difference becomes more pronounced at 115 K in the range $0.3<t_{delay}<2.5$ ps, then gradually diminishes upon further cooling. Importantly, the anisotropy observed at low temperature is an intrinsic property arising from symmetry breaking, not a pump-induced artifact, as confirmed by pump polarization rotation experiments (see Section II of Supporting Information). We note that previous ultrafast studies \cite{YMeng2024} employed a non-degenerate pump-probe scheme (400 nm / 800 nm), where high-energy excitations dominate the transient response. In contrast, our degenerate 800 nm pump-probe configuration primarily probes low-energy quasiparticle dynamics near the Fermi level, making it more sensitive to symmetry-breaking electronic responses such as nematicity.\setlength{\parskip}{0pt}

\begin{figure*}[t]
\vspace*{-0.2cm}
\begin{center}
\includegraphics[width=1.7\columnwidth]{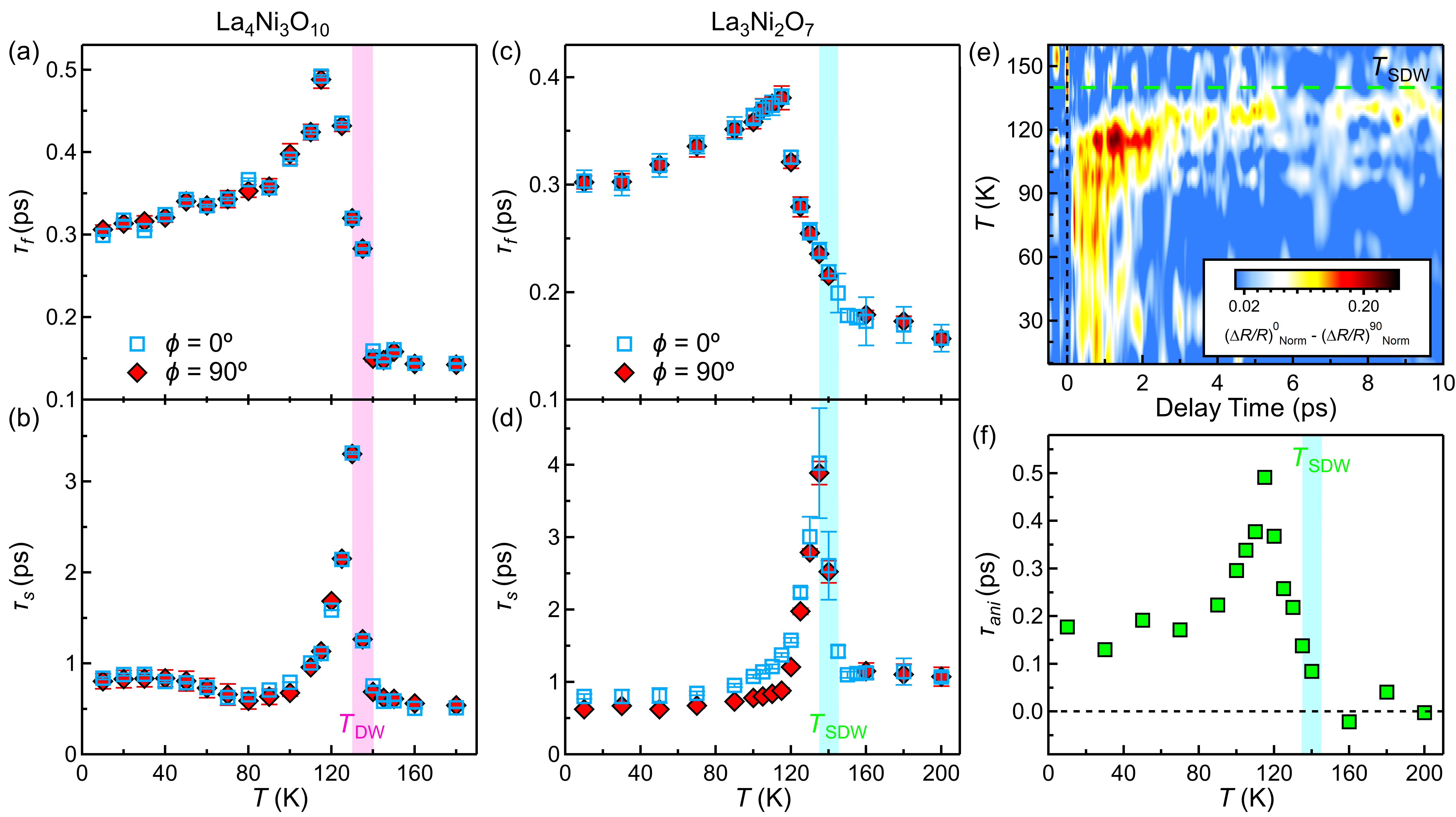}
\end{center}
\vspace*{-0.7cm}
\caption{(color online) The difference between $\phi$ = 0$^{\circ}$ and $\phi$ = 90$^{\circ}$ in La$_4$Ni$_3$O$_{10}$ and La$_3$Ni$_2$O$_7$. (a) and (b) Temperature dependence of the fast ($\tau_f$) and slow ($\tau_s$) relaxation times for La$_4$Ni$_3$O$_{10}$, respectively. (c) and (d) are same as (a) and (b) but for La$_3$Ni$_2$O$_7$. (e) The intensity difference between $\phi$ = 0$^{\circ}$ and $\phi$ = 90$^{\circ}$ [($\Delta R$/$R$)$_{ani}$ = $(\Delta R$/$R$)$^0_{\rm Norm} - (\Delta R$/$R$)$^{90}_{\rm Norm}$] for La$_3$Ni$_2$O$_7$. (f) Difference of the slow relaxation time $\tau_s$ between two polarizations, $\tau_{ani} = \tau^0_{s} - \tau^{90}_{s}$.}
\label{FIG:2}
\end{figure*}

To quantify these contrasting behaviors, we model the relaxation dynamics using a two-exponential decay (see Section III of Supporting Information) \cite{WTZhang2016, QYWu2025, CZhang2025}: a fast component $\tau_f$ associated with electron-boson thermalization and a slow component $\tau_s$ tracking quasiparticle (QP) relaxation across the density-wave gap. Figures \blue{2}(a)-\blue{2}(d) summarize the temperature dependence of $\tau_{f}$ and $\tau_{s}$ for both polarizations. Notably, in both compounds, the fast component lifetime $\tau_{f}$ exhibits a pronounced enhancement near the density-wave transition temperature. This reflects an acute sensitivity of the initial high-energy excitations and the subsequent thermalization process to the electronic reconstruction.

Focusing on La$_4$Ni$_3$O$_{10}$ first, $\tau_{f}$ and $\tau_{s}$ measured along both polarizations remain nearly coincident across the entire temperature range [Figures \blue{2}(a) and \blue{2}(b)]. Both components exhibit significant divergence near 135 K, representing a classic signature of the quasiparticle relaxation ``bottleneck'' upon gap opening \cite{ARothwarf1967, EChia2010, VVKabanov1999, JDemsar2006}. Fitting $\tau_{s}$ with the Rothwarf-Taylor (RT) model yields an isotropic zero-temperature gap $\Delta_{\rm DW}(0) \approx 26 \pm 6$ meV (see Section IV of Supporting Information), consistent with independent ARPES, Raman, and ultrafast spectroscopy measurements \cite{HLi2017, YLi2025, CZhang2025, SDeswal2024}. These observations establish $n$ = 3 {\LNO} as a conventional, isotropic density-wave system.

In contrast, bilayer La$_3$Ni$_2$O$_7$ shows a fundamentally different evolution. The fast relaxation times, $\tau^0_{f}$ and $\tau^{90}_{f}$, are nearly identical across all temperatures and show a significant downturn below 115 K [Figure \blue{2}(c)]. The onset of spontaneous symmetry breaking is most clearly manifested by the slow relaxation process, $\tau_s$, where the components measured along two polarizations diverge significantly below 140 K due to the opening of the SDW gap [Figure \blue{2}(d)]. We emphasize that $\tau^{0}_{f, s}$ and $\tau^{90}_{f, s}$ serve as effective parameters reflecting the anisotropic projection of intertwined relaxation channels onto the probe polarization, rather than representing inherently polarization-dependent intrinsic lifetimes. The RT model fitting provides polarization-dependent effective bottleneck energy scales: $\Delta_{\rm SDW}^0(0) \sim$ 69 $\pm$ 4 meV \cite{QYWu2025} and $\Delta_{\rm SDW}^{90}(0) \sim$ 54 $\pm$ 5 meV  (see Section IV of Supporting Information). It should be noted that this two-gap RT analysis is phenomenological, aimed at capturing the multiple effective energy scales arising from the complex, multi-orbital Fermi surface of La$_3$Ni$_2$O$_7$; rather than implying decoupled electronic subsystems, these extracted scales reflect the projection of an anisotropic gap structure \cite{GHe2025} onto the observed relaxation pathways. This anisotropy is quantified by the difference signal ($\Delta R$/$R$)$_{ani}$ [Figure \blue{2}(e)] and the dynamic anisotropy $\tau_{ani} = \tau^0_{s} - \tau^{90}_{s}$ [Figure \blue{2}(f)]. Below $T_{\rm SDW}$, $\tau_{ani}$ increases and peaks sharply near 115 K, and then decreases, confirming that the anisotropy induced by the broken symmetry develops and evolves within the SDW phase. While equilibrium probes have suggested anisotropic correlations in La$_3$Ni$_2$O$_{7}$ \cite{NKGupta2025, GHe2025}, our time-resolved measurements demonstrate that this symmetry breaking directly governs the non-equilibrium quasiparticle recombination bottleneck. Moreover, the non-monotonic temperature dependence of $\tau_{ani}$ reveals a dynamical signature of intertwined electronic orders, suggesting coupling or competition between nematicity and a secondary low-temperature DW-like instability, rather than establishing $\tau_{ani}$ itself as an independent thermodynamic order parameter.

\begin{figure*}[t]
\vspace*{-0.2cm}
\begin{center}
\includegraphics[width=1.9\columnwidth]{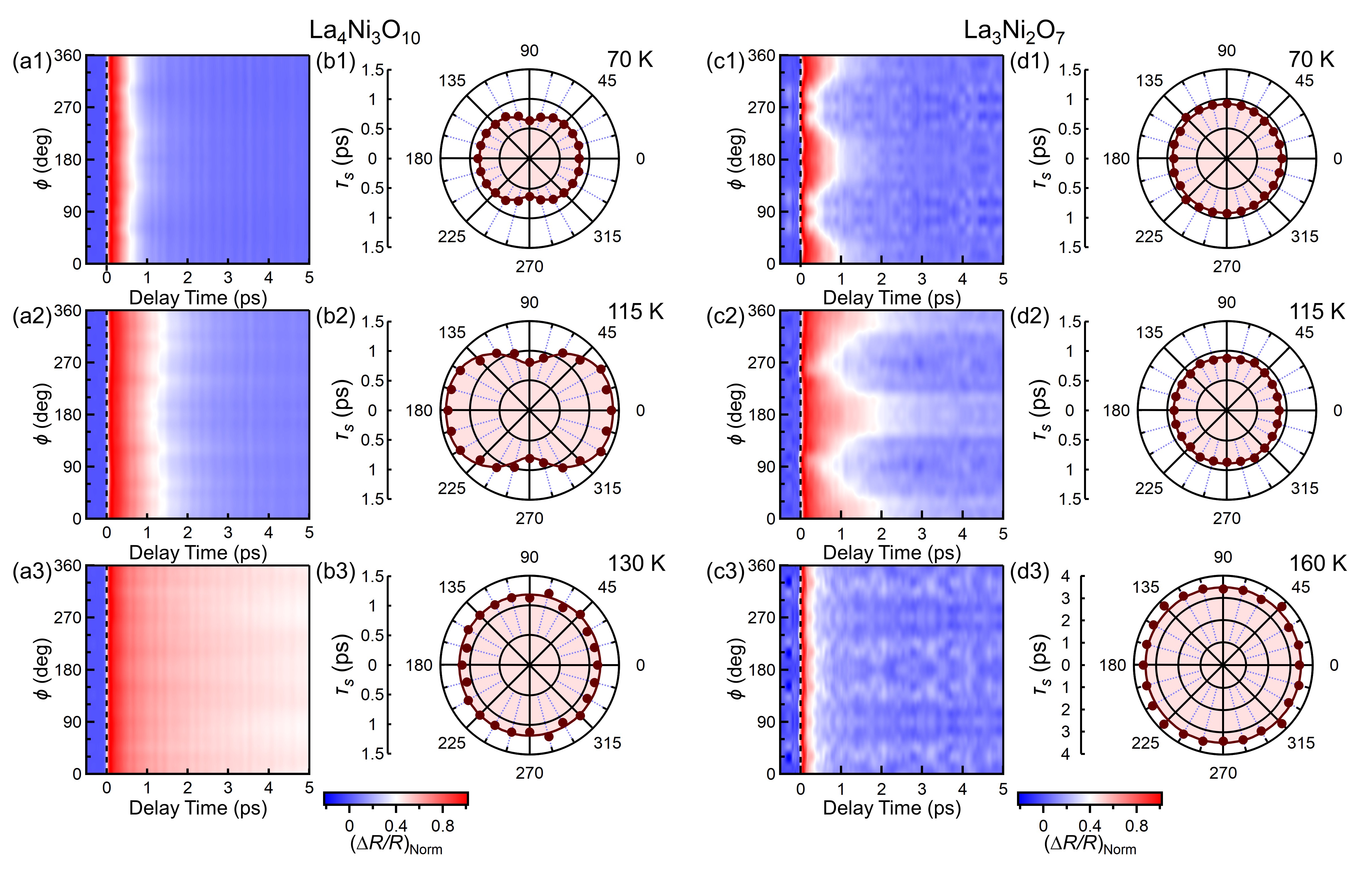}
\end{center}
\vspace*{-0.7cm}
\caption{(color online) Polarization dependence of normalized transient reflectivty and relaxation Time in  La$_4$Ni$_3$O$_{10}$ and La$_3$Ni$_2$O$_7$. (a) and (c) The angular dependence of the normalized transient reflectivity, ($\Delta R(\theta)$/$R$)$_{\rm Norm}$, measured at typical temperatures for La$_4$Ni$_3$O$_{10}$ and La$_3$Ni$_2$O$_7$ samples, respectively. (b) and (d) The polar plots of the relaxation time $\tau_s$, obtained by fitting the data with the two-exponential decay model. The solid lines in (b) and (d) are guide to eyes.}
\label{FIG:3}
\end{figure*}

To further characterize the symmetry of the dynamical response, we examined the angular dependence of ($\Delta R$/$R$)$_{\rm Norm}$ [Figures \blue{3}(a)-\blue{3}(d)]. For La$_4$Ni$_3$O$_{10}$, both ($\Delta R$/$R$)$_{\rm Norm}$ [Figure \blue{3}(a)] and the polar plot of $\tau_{s}$ [Figure \blue{3}(b)] remain circular and isotropic across all measured temperatures. Notably, while the maximum transient reflectivity amplitude exhibits a small anisotropy across all temperatures, consistent with the material's lower monoclinic crystal symmetry, the normalized relaxation dynamics ($\tau_{s}$) remain isotropic. This absence of dynamic anisotropy signifies that the trilayer compound maintains its electronic rotational symmetry despite the underlying lattice structure. It also demonstrates that the formation of a DW bottleneck alone does not necessarily generate a $C_2$-symmetric ultrafast response under our experimental geometry. Conversely, La$_3$Ni$_2$O$_7$ exhibits pronounced rotational symmetry breaking. At high temperature (160 K), $\tau_{s}$ is isotropic [Figures \blue{3}(c1) and \blue{3}(d1)], but below $T_{\rm SDW}$, a clear twofold ($C_2$) rotational symmetry emerges [Figures \blue{3}(c2) and \blue{3}(d2)], persisting down to 70 K [Figures \blue{3}(c3) and \blue{3}(d3)]. The emergence of a pronounced $C_2$ symmetry in $\tau_{s}$ only below $T_{\rm SDW}$, occurring without a reported structural transition  \cite{HWang2025}, identifies the observed state as a manifestation of intrinsic electronic nematicity, or more specifically, as a dynamical manifestation of electronic nematicity in the SDW state, rather than a secondary consequence of lattice distortion.

This electronic anisotropy is likely rooted in the magnetic degrees of freedom. Resonant soft X-ray studies suggest the SDW in La$_3$Ni$_2$O$_7$ forms a unidirectional diagonal spin stripe, creating nanoscale anisotropic domains \cite{NKGupta2025}. The resulting difference in correlation lengths reflects rotational symmetry breaking at the nanoscale \cite{NKGupta2025}. Furthermore, electronic Raman scattering also reports anisotropic electronic correlations in the SDW state \cite{GHe2025}. Also the anisotropy may come along with a possible twinning effect \cite{YNZhang2024}. We note that the recent studies using scattering-type scanning near-field optical microscopy have observed a stripy optical conductivity pattern in La$_3$Ni$_2$O$_7$, with adjacent domains oriented perpendicularly to each other \cite{XZhou2024}. Therefore, the anisotropy observed in this study might be contributed by a single domain, given the small probe spot employed. The observation of macroscopic anisotropy below $T_{\rm SDW}$  suggests that the probe spot preferentially samples a dominant electronic domain, or that the electronic nematicity is sufficiently robust to be resolved despite the underlying twinning typically found in these nickelates.

The non-monotonic behavior of $\tau_{ani}$ (peaking near 115 K and weakening at lower $T$ [Figure \blue{2}(f)] provides a dynamical indication of intertwined electronic orders. The nematic anisotropy, which is observed below $T_{\rm SDW}$, appears suppressed by a secondary order with $C_{4}$ symmetry emerging below $T^* \sim$ 115 K. Recent RXS experiments observed a charge-origin anomaly at 110 K \cite{XRen2025}. Additionally, the optical measurements revealed the opening of an energy gap below 115 K, which may indicate the formation of CDW order \cite{ZLiu2024, GJiang2025}. Therefore, the order  with $C_{4}$ symmetry emerging below 115 K might be associated with CDW order. As this isotropic order parameter grows upon further cooling, it progressively diminishes the $C_{2}$ nematic signature in the QP relaxation dynamics, leading to the observed reduction in $\tau_{ani}$.

The present measurements reveal a clear breaking of in-plane rotational symmetry in the bilayer nickelate La$_3$Ni$_2$O$_7$, whereas no such behavior is detected in the trilayer La$_4$Ni$_3$O$_{10}$. The emergence of a two-fold electronic response upon cooling recalls the residual nematic phase observed in iron-based superconductors \cite{SLiu2018, HHKuo2016, EThewalt2018}, as well as analogous behaviors in cuprates \cite{JWu2017, YToda2014}, kagome metals \cite{YXiang2021, LNie2022}, and heavy-fermion systems \cite{FRonning2017, ROkazaki2011}. In La$_3$Ni$_2$O$_7$, the electronic anisotropy is observed below the SDW transition, and manifests exclusively in the slow relaxation channel, which is directly coupled to the spin degrees of freedom \cite{QYWu2025}. These observations naturally implicate spin-driven symmetry breaking: the opening of a spin-density-wave (SDW) gap modifies quasiparticle relaxation pathways differently along the crystallographic $a$ and $b$ axes, producing a distinct $C_4 \rightarrow C_2$ response. Furthermore, we clarify that while the RT model describes the isotropic quasiparticle relaxation bottleneck, the observed electronic anisotropy primarily stems from the anisotropic optical matrix elements. This distinction ensures that the polarization-dependent dynamics correctly reflect the underlying symmetry breaking, even when interpreted through a simplified bottleneck framework. An orbital mechanism remains possible as well. In nickelate Ruddlesden-Popper phases, both experiment and theory indicate that the low-energy electronic structure is dominated by Ni-3$d$ states \cite{VChristiansson2023, FLechermann2023, YZhang2023, JYang2024, JLi2024, HLi2017}, and the optical response at 1.55 eV primarily probes Ni 3$d$-3$d$ transitions. If orbital-selective nematicity develops, unequal orbital occupations along the two in-plane axes would give rise to the polarization-dependent probe response observed here.

The absence of such anisotropy in La$_4$Ni$_3$O$_{10}$ is notable. Compared with the bilayer, the trilayer hosts reduced electronic correlations, a more intricate band structure, and, as demonstrated, a layer-dependent pattern of intertwined density waves in which the SDW node resides on the inner Ni-O plane while the outer planes exhibit an out-of-phase modulation \cite{JZhang2020}. This structural and electronic complexity likely suppresses the formation of a uniform in-plane nematic susceptibility, thereby explaining the isotropic ultrafast response.

Nematicity has emerged as a unifying theme across several families of unconventional superconductors. In both cuprates and iron-based systems, extensive evidence points to a nematic quantum critical point (QCP) near optimal doping, where the superconducting transition temperature $T_c$ is maximal \cite{KFujita2014, KIshida2020, JHChu2010, PWalmsley2013}. Near such a QCP, the amplitude and correlation length of nematic fluctuations diverge, producing long-range interactions that can destabilize conventional metallic behavior and promote either non-Fermi-liquid transport or superconductivity \cite{WMetzner2003, AEBohmer2022}. Recent theoretical analyses further suggest that nematic fluctuations generate an attractive effective interaction across multiple pairing channels, allowing them either to enhance a pairing state driven by other interactions or, in some scenarios, to serve as the dominant pairing glue  \cite{AEBohmer2022, SLederer2015, TAMaier2014, SLederer2017}.

The rotational-symmetry breaking identified in La$_3$Ni$_2$O$_7$ below $T_{\rm SDW}$, and its absence in La$_4$Ni$_3$O$_{10}$, provide a materials-based link between nematicity and superconductivity in the nickelates. This contrast likely originates from orbital-selective physics: strong $d_{z^2}$ hybridization across Ni-O-Ni vertical dimers in La$_3$Ni$_2$O$_7$ promotes a flat electronic structure that is susceptible to nematic fluctuations. Specifically, the vertical Ni-O-Ni linkage enhances interlayer $d_{z^2}$ hopping and produces bonding-antibonding splitting, giving rise to relatively flat $d_{z^2}$-derived electronic states near the Fermi level. Such flat electronic structures enhance the low-energy density of states and electronic susceptibility, thereby amplifying orbital-selective or spin-stripe-driven nematic fluctuations in the multi-orbital background. In La$_4$Ni$_3$O$_{10}$, the additional NiO$_2$ layer does not simply enhance the same bilayer-like coupling; instead, it redistributes the $d_{z^2}$-derived electronic states over inequivalent inner and outer NiO$_2$ planes, preserving  the original $C_4$ symmetry and suppressing anisotropy. This layer differentiation is consistent with the layer-dependent intertwined density-wave order reported in the trilayer compound and may frustrate the formation of a uniform in-plane nematic susceptibility. If such nematic fluctuations provide an additional pairing interaction alongside magnetic fluctuations, they may contribute to the enhanced $T_c$ observed in bilayer nickelates under pressure. These results therefore highlight nematicity as a potentially important ingredient in the pairing mechanism of layered nickelates. To verify this idea, we expect future experiments on the pressure response of nematic order parameter in both La$_3$Ni$_2$O$_7$ and La$_4$Ni$_3$O$_{10}$ to directly probe its relation with superconductivity. Furthermore, considering recent debates on whether bulk superconductivity follows the structural transition from orthorhombic to tetragonal symmetry \cite{ChenXianHui2025, Wangmeng2511, ZHuo2025B}, we anticipate that electronic nematicity, often associated with crystal deformation and twinning, will play a central role in uncovering the complex interplay among structural transitions, density waves, and high-$T_c$ pairing.

In summary, our work provides a direct contrast of the normal-state electronic symmetries in bilayer La$_3$Ni$_2$O$_7$ and trilayer La$_4$Ni$_3$O$_{10}$ nickelates using polarized ultrafast spectroscopy. While both materials exhibit density-wave (DW) transitions, the trilayer La$_4$Ni$_3$O$_{10}$ maintains a robustly isotropic optical response over the entire temperature range. In marked distinction, La$_3$Ni$_2$O$_7$ displays a pronounced rotational-symmetry-breaking electronic response with twofold anisotropy. This electronic nematicity manifests as a significant anisotropy in the slow quasiparticle relaxation dynamics and evolves within the temperature regime where the DW gap is well developed. Crucially, the non-monotonic temperature dependence of the nematic signal indicates its coupling or competition with a secondary, isotropic DW-like order below 115 K. The striking contrast in electronic symmetry between the bilayer and trilayer systems highlights the potential relevance of symmetry-breaking electronic correlations in bilayer nickelates and motivates further investigation into their possible connection to superconductivity under pressure.

This work was supported by the National Natural Science Foundation of China (Grants No. 92265101, No. 12074436, No. 12425404, No. 12494591, No. 92565303), the National Key Research and Development Program of China (Grants No. 2022YFA1604204, No. 2023YFA1406500, No. 2022YFA1402802), the Open Project of Beijing National Laboratory for Condensed Matter Physics (Grant Nos. 2024BNLCMPKF001), the Science and Technology Innovation Program of Hunan province (2022RC3068), the Guangdong Basic and Applied Basic Research Funds (Grant No. 2024B1515020040), Guangdong Major Project of Basic Research (2025B0303000004) ,Guangdong Provincial Key Laboratory of Magnetoelectric Physics and Devices (Grant No. 2022B1212010008), Research Center for Magnetoelectric Physics of Guangdong Province (2024B0303390001), and Guangdong Provincial Quantum Science Strategic Initiative (Grant No. GDZX2401010).

\end{document}